\documentclass[twocolumn,pra,aps]{revtex4}
\usepackage{graphicx}

\begin{document}

\title{Why aren't quantum correlations maximally nonlocal?\\Biased local randomness as essential feature of quantum mechanics.}

\author{Antoine Suarez}
\address{Center for Quantum Philosophy, P.O. Box 304, CH-8044 Zurich, Switzerland\\
suarez@leman.ch, www.quantumphil.org}

\date{February 14, 2009}

\begin{abstract} It is argued that the quantum correlations are not maximally nonlocal to make it possible to control local outcomes from outside spacetime, and quantum mechanics emerges from timeless nonlocality and biased local randomness. This rules out a world described by NL (nonlocal) boxes. A new type of experiments is suggested.

\end{abstract}

\pacs{03.65.Ta, 03.65.Ud, 03.30.+p, 04.00.00, 03.67.-a}

\maketitle

Bell type experiments demonstrate (within the limits of a few rather eccentric loopholes) correlations, which cannot be explained by means of local relativistic influences propagating at velocity $v \leq c$ \cite{jb}. This means that one has to give up the view that the outcomes at each part of the setup result from properties preexisting in the particles before measurement: Alice's (respectively Bob's) outcomes cannot be explained by the information the photon carries when leaving the source and the settings in Alice's (respectively Bob's) lab.

The Suarez-Scarani or \emph{before-before} experiment demonstrates that these nonlocal correlations cannot be explained by time-ordered nonlocal influences \cite{asvs97,as00.1,szsg}. Giving up the concept of locality is not sufficient to be consistent with quantum experiments, one has to give up the view that one event occurring before in time can be considered the cause, and the other occurring later in time the effect (nonlocal determinism). The correlations cannot be explained by any history in spacetime, in entanglement experiments local random events experience influences from outside spacetime to produce nonlocal order. \cite{as08,cb}

However, the orthodox interpretation of quantum mechanics also claims that in entanglement experiments the local outcomes happen in a ``full random'' way, i.e., according to a uniform (non-biased) random distribution. In this sense, the orthodox interpretation is at variance with any model assuming that local outcomes can happen according to a biased random distribution. Both Bell's and Suarez-Scarani's experiments are compatible with such models.

The violation of Leggett inequalities was first interpreted as an experimental falsification of ``nonlocal realism'', where ``realism'' refers to the view that the single particles carry well defined properties when they leave the source \cite{gro}. Such an interpretation is misleading: By testing models fulfilling Leggett inequalities one does not test ``nonlocal realism'', but rather models assuming both nonlocal randomness and outcomes that depend on biased random local variables \cite{as09}. Nevertheless, it is the Colbeck-Renner theorem \cite{core} which clearly shows the relationship between nonlocality and biased local randomness in entanglement experiments \cite{as09}.

In this letter I argue that the quantum correlations are not maximally nonlocal to make it possible biasing local outcomes from outside spacetime, and propose to consider timeless nonlocality and biased local randomness as primitives to axiomatize quantum theory. This rules out a world described by NL boxes. I also propose a new type of entanglement experiments demonstrating these ideas.

\begin{figure}[t]
\includegraphics[width=80 mm]{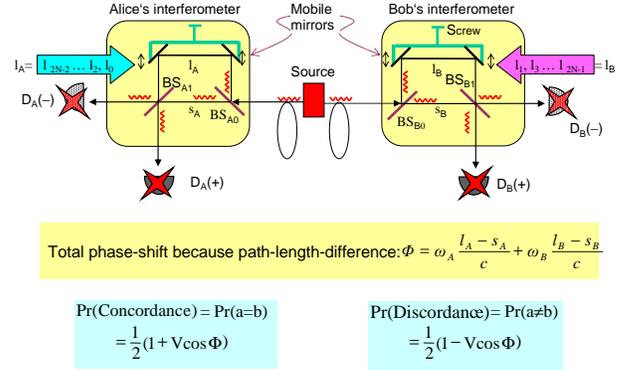}
\caption{Diagram of a chained Bell experiment using interferometers. The setup makes it possible to perform at the same time a before-before experiment using acousto-optic waves as moving beam-splitters. (See text for details).}
\label{f1}
\end{figure}

Consider the experiment sketched in Figure \ref{f1}: The source emits photon pairs. Photon A (frequency $\omega_{A}$) enters Alice's interferometer to the left through the beam-splitter BS$_{A0}$ and gets detected after leaving the beam-splitter BS$_{A1}$, and photon B (frequency $\omega_{B}$) enters Bob's interferometer to the right through the beam-splitter BS$_{B0}$ and gets detected after leaving the beam-splitter BS$_{B1}$. The detectors are denoted D$_{A}(a)$ and D$_{B}(b)$ ($a, b\in\{+,-\}$). Each interferometer consists in a long arm of length $l_{i}$, and a short one of length $s_{i}$, $i\in\{A,B\}$. Frequency bandwidths and path alignments are chosen so that only the coincidence detections corresponding to the path pairs: $(s_{A},s_{B})$ and $(l_{A},l_{B})$ contribute constructively to the correlated outcomes in regions A and B, where $(s_{A},s_{B})$ denotes the pair of the two short arms, and $(l_{A},l_{B})$ the pair of the two long arms.

Suppose one of the measurements produces the value $a$ ($a\in\{+,-\}$), and the other the value $b$ ($b\in\{+,-\}$).
According to quantum mechanics the probability $Pr(a, b)$ of getting the joint outcome $(a, b)$ depends on the choice of the phase parameter $\Phi$ characterizing the paths or channels uniting the source and the detectors:

\begin{eqnarray}\label{1}
    Pr(a=b)=\frac{1}{2}(1+V\cos\Phi) \nonumber\\
    Pr(a\neq b)=\frac{1}{2}(1-V\cos\Phi)
\end{eqnarray}

\noindent where $\Phi$ is the phase parameter given by: $\Phi=\omega_{A}\frac{l_A - s_A}{c}+\omega_{B}\frac{l_B - s_B}{c}$.

Bell experiments, using two different values of $l_{A}$ and two different values of $l_{B}$, demonstrate that the correlations violate locality criteria, the well known Bell's inequalities (see \cite{szsg} and references therein).

In the before-before experiment the beam-splitters BS$_{A1}$ and BS$_{B1}$ are in motion in such a way that each of them, in its own reference frame, is first to select the output of the photons (before-before timing). Then, each outcome should become independent of the other, and (according to a time-ordered causal model) the nonlocal correlations should disappear. The result was that the correlations don't disappear, and therefore are independent of any time-order \cite{szsg}, that is, they come from outside spacetime \cite{as08,cb}.

Consider now chained Bell experiments using $N$ different values of $l_{A}$ ($l_0, l_2,...,l_{2N-2}$) and $N$ values of $l_{B}$ ($l_1, l_3,...,l_{2N-1}$), with $N\geq2$. We define the function $I(N)$ as:
\begin{eqnarray}\label{2}
    I(N)&=& Pr(a=b|\Phi(l_{0},l_{2N-1})) \nonumber\\
    &+& Pr(a\neq b|\Phi(l_{0},l_{1})) \nonumber\\
    &+& Pr(a\neq b|\Phi(l_{1},l_{2})) \nonumber\\
    &+& ....... \nonumber\\
    &+& Pr(a\neq b|\Phi(l_{2N-2},l_{2N-1}))
\end{eqnarray}

\noindent where $Pr(a\!=\!b|\Phi(l_{0},l_{2N-1}))$ means the conditional probability that Alice and Bob get the same outcome if the phase's value results from long interferometers' arms set to $l_{0},l_{2N-1}$, and $Pr(a\!\neq\! b|\Phi(l_{i},l_{i+1}))$ the conditional probability that Alice and Bob get different outcomes if the phase's value results from long interferometers' arms set to $l_{i},l_{i+1}$; depending on $i$, $l_{i}$ denotes the arm of Alice's or Bob's interferometer.

\begin{figure}[t]
\includegraphics[width=80 mm]{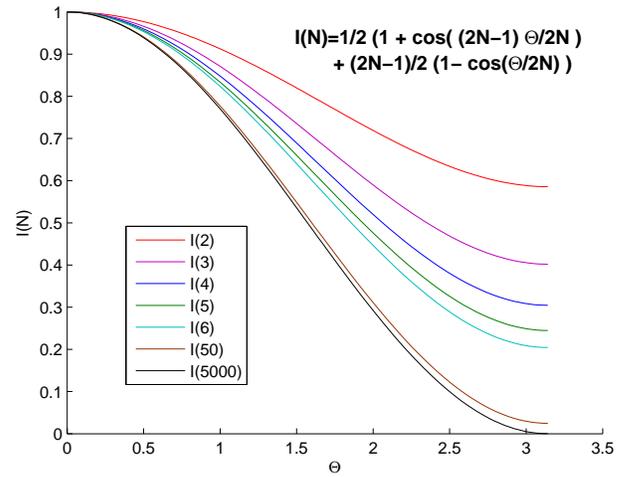}
\caption{$I(N)$ as function of $\Theta$, for different values of $N$ (see Equation (\ref{5}) in text).}
\label{f2}
\end{figure}

We assume that any two values $l_{i}, l_{i+1}$, with $i\in\{0, 2N-2\}$, define the same phase parameter, resulting from the equipartition of a value $\Theta$:
\begin{equation}\label{3}
\Phi(l_{i}, l_{i+1})=\Theta/2N
\end{equation}

Then, equation (\ref{2}) can be rewritten as follows:
\begin{eqnarray}\label{4}
    I(N)&=& Pr\left(a=b|(2N-1)\frac{\Theta}{2N}\right)\nonumber\\
    &+& (2N-1)Pr\left(a\neq b|\frac{\Theta}{2N}\right)
\end{eqnarray}

For each $N$, $I(N)\geq1$ defines a Bell inequality or locality criterion. $I(2)\geq1$ represents the well known CHSH inequality for experiments with 4 measurements. Accordingly, $I(N)<1$ defines correlations that cannot be explained by means of local relativistic influences, and decreasing $I(N)$ can be considered an indicator of increasing nonlocality.

We denote $D(N)$ the statistical distance between the distribution of the local outcomes and the uniform random distribution in the corresponding chained Bell experiment with $2N$ measurements.

The Colbeck-Renner \cite{core} theorem establishes that:
\begin{equation}\label{13}
D(N)\leq I(N)/2
\end{equation}

Substituting (\ref{1}) into (\ref{4}) gives:

\begin{eqnarray}\label{5}
    I(N)&=& \frac{1}{2}\left(1-cos\left((2N-1)\frac{\Theta}{2N}\right)\right)\nonumber\\
    &+& \frac{2N-1}{2}\left(1-cos\frac{\Theta}{2N}\right)
\end{eqnarray}

Figure \ref{f2} represents $I(N)$ in function of the phase parameter $\Theta$, for different $N$. As it appears, the plotted quantum mechanical prediction is not consistent with the ``orthodox interpretation'' that the local outcomes are uniformly distributed for any entanglement experiment. The fact that $I(2)=2-\sqrt{2}>0$ for $\Theta=\pi$ clearly suggests that Nature is keen to permit biased local random outcomes for any value of the phase parameter $\Theta$.

Now I prove theorems showing that essential features of quantum mechanics emerge from nonlocality and biased local randomness.

\emph{Basic conditions:} For reasons of scaling and symmetry we impose:
\begin{eqnarray} \label{8}
&&Pr(a=b|\Phi=0)=1\\\label{9}
&&Pr(a=b|\Phi)=Pr(a\neq b|\pi-\Phi)
\end{eqnarray}

Additionally, we take account of the fact that nature likes ``smoothness'' for fashioning distributions and assume:
\begin{eqnarray}
\label{10}
&&I(N,\Theta_1)>I(N,\Theta_2)=0,\; \Theta_1<\Theta<\Theta_2\nonumber\\
&&\Rightarrow I(N,\Theta_1)>I(N,\Theta)>0\\
\label{11}\nonumber\\
&&1>I(2,\pi)>I(\infty,\pi)\nonumber\\
&&\Rightarrow I(2,\pi)>I(N,\pi)>I(N+1,\pi)>I(\infty,\pi),\nonumber\\
&&\forall N>2
\end{eqnarray}

\noindent where $I(N,\Theta)$ denotes the value of the function $I(N)$ for the phase parameter $\Theta$.\\

\emph{Theorem 1:} $I(2,\pi)\!=0\! \Rightarrow 1\!>\!I(2,\Theta)\!>\!0$ for $ 0<\!\Theta\!<\!\pi$. That is, maximal nonlocality for $\Theta\!=\!\pi$ necessarily implies non-maximal nonlocality for $0<\Theta<\pi$.\\

\emph{Proof:} From Equations (\ref{4}) and (\ref{8}) one is led to:

\begin{eqnarray}\label{6}
   I(N, \Theta=0)&=&Pr(a=b|\Phi=0)\nonumber\\
   &+&2N Pr(a\neq b|\Phi=0)\nonumber\\
    &-& Pr(a\neq b|\Phi=0)=1
\end{eqnarray}

Then, from the ``smoothness'' condition (\ref{10}) the \emph{Theorem 1} follows.\\

\emph{Theorem 2: } $1>I(2,\pi)>0\Rightarrow \forall N>2,\; I(2,\pi)>I(N,\pi)>I(N+1,\pi)>I(\infty,\pi)=0$. That is, non-maximal nonlocality for $\Theta=\pi$ and $N=2$ is a sufficient condition for decreasing $I(N,\pi)$:\\

\emph{Proof:} Taking account of (\ref{9}), Equation (\ref{4}) implies for $N=\infty$:
\begin{eqnarray}\label{7}
   I(\infty, \Theta=\pi)&=&Pr(a=b|\Phi=\pi)\nonumber\\
   &+&2N (1-Pr(a=b|\Phi=0)\nonumber\\
    &-& Pr(a\neq b|\Phi=0)=0
\end{eqnarray}

\begin{figure}[t]
\includegraphics[width=80 mm]{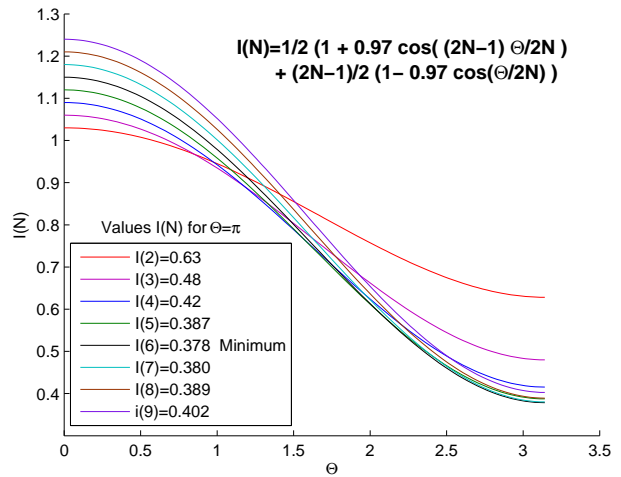}
\caption{Functions $I(N)$ with Visibility $V=0.97$. For $\Theta=\pi$, the values $I(N)$ exhibit a minimum at I(6)}
\label{f3}
\end{figure}

Then, from the ``smoothness'' condition (\ref{11}) the \emph{Theorem 2} follows.

\emph{Theorems 1 and 2} mean that making it possible to bias local random outcomes is a necessary and sufficient condition to get nonlocal distributions fashioned like those represented in Figure \ref{f2}. In particular, nonlocal nature cannot be maximally nonlocal for all phases.

This rules out a world described by NL (nonlocal) boxes. Actually, one would need a whole spectrum of NL boxes with biases ranging from $I(\infty,\pi)=0$ to $I(2,0)=1$.

However a question remains open: Why the particular Bell value $I(2,\pi)=2-\sqrt{2}$, instead of for instance $I(2,\pi)=2-\sqrt{3}$? Is it simply motivated by the wish of choosing ``nice enough'' functions like in (\ref{1}), in order to make the work more enjoyable to the physicists, or is there a deeper reason behind?

In conclusion, the preceding analysis shows that the quantum correlations are not maximally nonlocal to make it possible to bias local outcomes from outside spacetime. Thus, nonlocality without signaling and biased local randomness have strong primitive-appeal to explaining why the laws of nature are quantum. This means that entanglement experiments demonstrating nonlocality alone are basically incomplete, and should be expanded to experiments demonstrating nonlocality, timelessness and increasingly uniform bias altogether.

Assuming a visibility factor $V (0\leq V\leq1)$ depending mainly on the efficiency of the detectors, Equation (\ref{5}) becomes:
\begin{eqnarray}\label{12}
    I(N)&=& \frac{1}{2}\left(1-Vcos\left((2N-1)\frac{\Theta}{2N}\right)\right)\nonumber\\
    &+& \frac{2N-1}{2}\left(1-Vcos\frac{\Theta}{2N}\right)
\end{eqnarray}

Experiments with visibility $V=0.99$ are possible using resting beam-splitters \cite{hw}, and with $V=0.97$ using beam-splitters in motion (i.e. acousto-optic modulators) \cite{szsg}. This means, according to (\ref{13}) and the values of $I(N)$ in (\ref{12}) represented in Figure \ref{f3}, that a before-before experiment demonstrating a bias bound decreasing from $D=I(2)/2=0.315$ to $D=I(2)/2=0.189$ is feasible.

I would like to finish by stressing that the possibility of controlling outcomes from outside spacetime has a very natural correlate in the way the brain functions. When I am typewriting this article, I assume that the author is the same who typewrote the article proposing the before-before experiment in 1997. In this sense my identity has roots beyond spacetime. I am controlling the outcomes of my brain, i.e. biasing the random firings of my neurons, from outside spacetime. The result presented in this paper upholds the view that quantum randomness does not exclude the possibility of order and control and, therefore, remains susceptible of being influenced by free will \cite{su08}.\\
\\
\noindent\emph{Acknowledgments}: I am grateful to Roger Colbeck for insightful discussions, and acknowledge information in view of experiments by Nicolas Gisin, Renato Renner, Harald Weinfurter, and Hugo Zbinden.

\end{document}